\documentclass[12pt]{article}
\usepackage{epsfig,rotating,latexsym}
\begin{document}
 \begin{titlepage}
 \parindent 0pt
 \font\big=cmbx10 scaled\magstep2
\vskip .2cm
 May, 1997     \hfill\break
\vskip 1cm
 {\big\begin{center} 50 TeV HEGRA Sources and Infrared Radiation
 \end{center}}
\begin{center}
\vskip .3cm
V. Berezinsky\\
\vskip .3cm
 INFN, Laboratori Nazionali del Gran Sasso\\
67010 Assergi (AQ), Italy\\
\end{center}
\vskip 0.3cm
\begin{center}
Lars Bergstr\"om\\
Department of Physics, University of Stockholm\\
Box 6730, S--113 85 Stockholm, Sweden\\
\end{center}
\vskip0.3cm
\begin{center}
H. R. Rubinstein\\
 \vskip .3cm
Department of Theoretical Physics, Uppsala University\\
Box 803, S-751 08 Uppsala, Sweden\\
\end{center}
\vskip .2cm
\begin{abstract}
The recent observations of 50 TeV gamma radiation by HEGRA have
the potential of determining the  extragalactic flux of infrared 
radiation. The fact
that radiation is observed in the range between 30 and $100$~TeV
sets an  upper limit on the infrared flux, while  a cutoff 
at $E_{\gamma} \approx 50$~TeV fixes this flux with a  good 
accuracy.
If the intrinsic radiation is produced due 
to interaction of high energy protons with gas or low-energy target 
photons, then an  accompaning high-energy neutrino flux is unavoidable.
We calculate this flux and 
underground muon flux produced by it. The muon flux is dominated by muons
with energies about 1 TeV and can be marginally detected by a 1~km$^2$ 
detector like
an expanded AMANDA.

\end{abstract}
\vskip .7cm
\footnotesize\end{titlepage}

\newpage
\font\big=cmbx10 scaled\magstep1
\font\vbig=cmbx10 scaled\magstep2
\vskip .2cm
\noindent
\vskip .2cm
The detection by HEGRA \cite{hegra} of sources of gamma radiation 
with energy 
about 50 TeV, if confirmed, 
is a remarkable observation which hopefully is the first step
in ultra high energy gamma astronomy. Besides its own importance,
this 
radiation is an excellent tracer of infrared fluxes in intergalactic 
space. 

The existence of an upper limit
of the observed energies of photons is interpreted in \cite{hegra} 
as absorption on diffuse infrared radiation (IR). This assumption looks 
reasonable due to the following argument. Observation by CASA \cite{casa} 
has not resulted in detection of
these sources \cite{casa1}. This detector has much higher 
flux sensitivity than HEGRA, but also higher energy threshold 
($E_{th}=100$~TeV), which is sharp and makes ineffective the observations 
of gamma ray sources below about 50 TeV \cite{cronin}.
Therefore the cutoff 
of the spectrum at 50 TeV must be very sharp. The known acceleration 
mechanisms and in particular shock acceleration cannot provide such a 
sharp cutoff, while absorption naturally results in  an exponential 
cutoff. 

On the other hand, the observation of 50 TeV gamma-radiation from 
the HEGRA sources, which are typically at a distance greater than 
$200$~Mpc, implies that the flux of diffuse IR radiation is lower 
than was theoretically estimated 
(see \cite{Primack,others} and references 
therein\footnote{The high flux of IR radiation 
predicted theoretically has recently been 
questioned by Whipple observations \cite{trevor} of Mkn421 at energies of 
a few TeV.}).

It is interesting to note that recently the EASTOP 
collaboration analyzed their 
data on Mk 421, which was 
detected by HEGRA with $3.8 \sigma$ excess over background. At energy 
higher than $40$~TeV EASTOP collaboration established  \cite{navarra}
the $90\%$ c.l.
upper limit $1.2\cdot 10^{-13}$~cm$^{-2}$s$^{-1}$ close to the flux observed 
by HEGRA.
Within the statistical errors there is no serious contradiction 
between these two measurements. In our analysis we concentrate on the 
source 0116+319 detected by HEGRA at $5.7 \sigma$ level with flux 
$1.4 \cdot 10^{-13}$~cm$^{-2}$s$^{-1}$ at $E> 50$~TeV.

We shall discuss some consequences of the HEGRA observations.
One of them is the deriviation of the flux of IR radiation consistent 
with these observations.
If the intrinsic flux has its origin in the interaction of 
high energy protons with gas or target photons
then an accompanying high energy neutrino flux is unavoidable. 
We shall also discuss the detectability of this flux.

Let us parametrize the density of IR photons in intergalactic space, 
$n(\varepsilon)$, as
\begin{equation}
n(\varepsilon)=\frac{n_0}{\varepsilon_0} 
\left( \frac{\varepsilon}{\varepsilon_0} \right)^{-\nu},
\label{eq:den}
\end{equation}
where $\varepsilon$ is the energy of the IR photons and the
normalization  value is fixed 
at $\varepsilon_0=1\cdot 10^{-2}$~eV. One can formally use Eq.(\ref{eq:den}) 
for an
unlimited range of energies because at $\varepsilon \leq 3\cdot 10^{-3}$~eV
the microwave radiation dominates while at $\varepsilon \geq 0.3$~eV optical 
radiation does.

The probability of absorption (the inverse absorption length) of high energy 
photons with energy $E_{\gamma}$ is given by \cite{BBDGP}:
\begin{equation}
\frac{dW}{dl}=\frac{4}{E_{\gamma}^2}
\int_{m_e}^{\infty} d\varepsilon_c \sigma(\varepsilon_c)\varepsilon_c^3
\int_{\varepsilon_c^2/E_{\gamma}}^{\infty}d\varepsilon
\frac{n(\varepsilon)}{\varepsilon^2},
\label{eq:Wabs}
\end{equation}
where $m_e$ is the electron mass, $\varepsilon_c$ is the photon energy 
in the centre of momentum system 
of two colliding photons and $\sigma(\varepsilon_c)$ is the cross-section for 
pair production ($\gamma+\gamma \to e^++e^-$), which is given in terms 
of velocity $v=(1-m_e^2/\varepsilon_c^2)^{1/2}$ as \cite{Ber}:
\begin{equation}
\sigma(v)=\frac{3}{16}\sigma_T(1-v^2) \left[ (3-v^4)
\ln \left( \frac{1+v}{1-v}\right) +2v(v^2-2) \right],
\label{eq:crsec}
\end{equation}
where $\sigma_T$ is the Thompson cross-section.

After simple calculations one obtains:
\begin{equation}
l_{abs}^{-1}=\frac{dW}{dl}=\frac{3\Phi_{\nu}}{4(1+\nu)}n_0\sigma_T
\left( \frac{E_{\gamma}\varepsilon_0}{m_e^2} \right)^{\nu-1},
\label{eq:abs}
\end{equation}
where
\begin{equation}
\Phi_{\nu}=\int_{0}^{1}vdv(1-v^2)^{\nu-1}
\left[ (3-v^4)\ln \left( \frac{1+v}{1-v}\right) +2v(v^2-2) \right],
\label{eq:Phi}
\end{equation}
For integer values of $\nu$ this integral can be solved, e.g.,
\begin{equation}
\Phi_1=14/9,\,\,\,\Phi_2=22/45,\,\,\, \Phi_3=56/225.
\end{equation}
Over the whole range $1< \nu < 3$, a parametrization accurate to better
than one percent is given by
\begin{equation}
\Phi_\nu\approx 0.0791+1.857e^{-0.802\nu}+11.43e^{-2.876\nu}.
\end{equation}
For convenience, we tabulate $\Phi_\nu$ for some selected values of $\nu$
 in Table 1.
\begin{table}[htb]
\begin{center}
 \begin{tabular}{|c|c|c|c|c|c|c|c|c|c|}
 	\hline
 	$\nu$& 1.5 &1.7  &1.9  &2.0  &2.1 &2.3&2.5&2.7&2.9  \\
 	\hline
 $\Phi_\nu$ &0.789  &0.640  &0.532  &0.489  &0.451&0.387&0.337&0.297&0.263\\  
 	\hline
 \end{tabular}
\end{center}
 \label{tab:phi}
 \caption{The value of the function $\Phi_{\nu}$ of 
 Eq.\,~(\ref{eq:Phi}) for some values of $\nu$.}
\end{table}

Now one can put an upper limit on the density of IR photons for the energy 
range $10-50$~TeV (the range of HEGRA detectability for the set of detectors
used in the analysis in \cite{hegra}) imposing the condition 
$l_{abs} \geq r_s$, where $r_s=(c/H_0)z$ is a distance to a HEGRA 
source\footnote{Since all three HEGRA sources have redshifts $z<0.06$, the 
effects of cosmological evolution are very small.} 
(we use the Hubble constant 
$H_0=75$~km\,s$^{-1}$Mpc$^{-1}$).

\begin{figure}[ht]
\includegraphics[width=15cm]{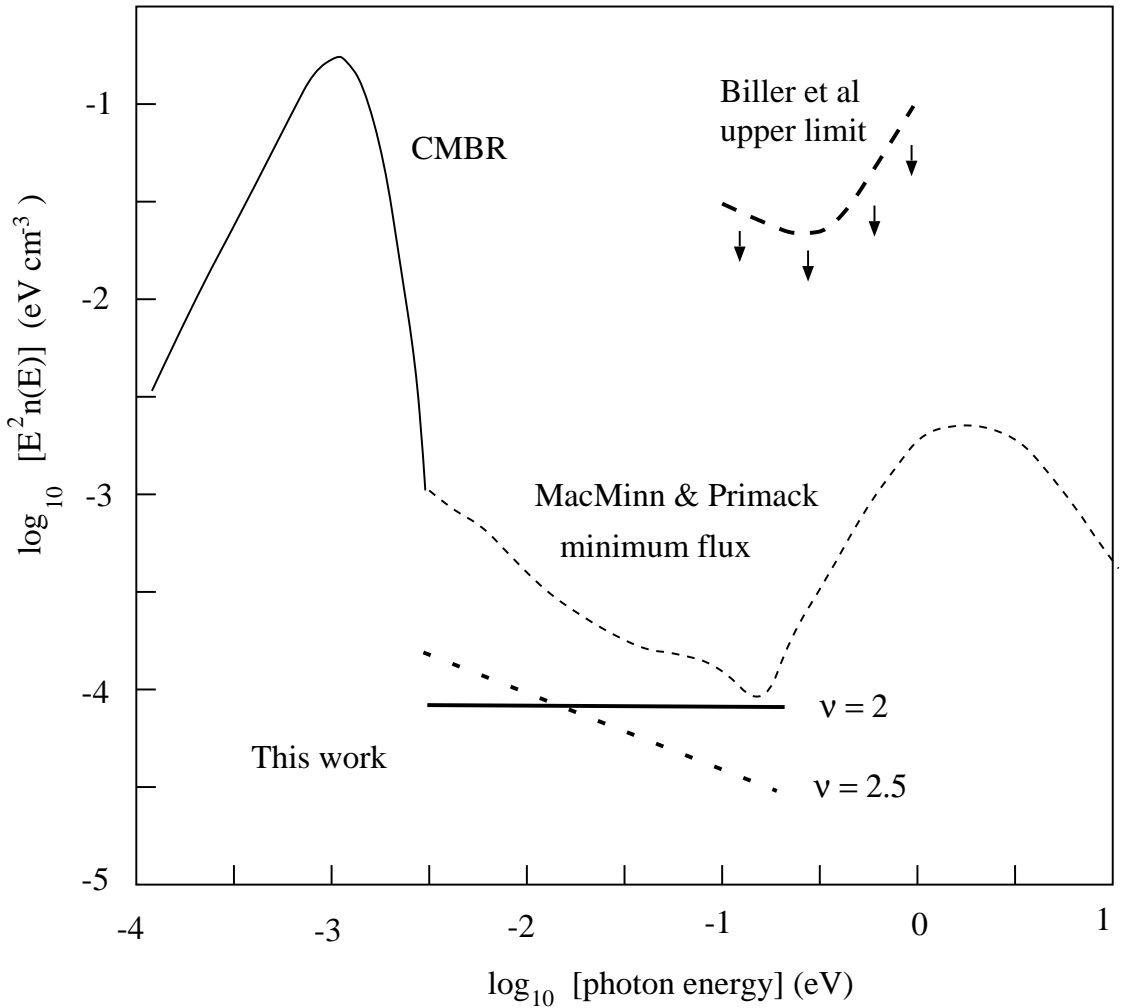}
\caption{Energy density of the extragalactic radiation field as a 
function of photon energy (logarithmic scales). The curves shown are 
the cosmic microwave background (CMBR), the lowest curve of the 
analysis of MacMinn and Primack [4], the upper limit from the 
analysis of Biller et al. [10], and  the flux needed for interpretation 
of HEGRA data according to our analysis for two values of the slope 
parameter $\nu$ of the IR flux, $\nu=2$ and $\nu=2.5$.}
\label{fig:fig1}
\end{figure}

To explain the HEGRA results we have to assume $l_{abs} \approx r_s$
at $E_{\gamma} \approx 50$~TeV. This condition fixes the flux of IR 
radiation for a given $\nu$. The calculated fluxes are exposed in 
Fig.\,\ref{fig:fig1} for 
different values of $\nu$. One can see that these fluxes are 
much lower than ones estimated before.  

The probability of photon absorption (inverse absorption length) is 
displayed in Fig.\,~2 together with inverse distance to the source 
0116+39 (redshift z=0.059) shown by the horizontal line. One can observe 
the dramatic increase of absorption at $E_{\gamma}=100$~TeV. The 
differential flux is suppressed by a factor of $60-70$ in comparison 
with that at $50$~TeV, and integral spectrum is suppressed even more
due to the sharp cutoff of spectrum at $100$~TeV. This may explain why 
the source  is not observed by the CASA detector.

Photons of the intrinsic radiation with energy between 50 and 100 TeV 
are absorbed on IR radiation and photons with higher energies  on 
microwave radiation. The produced electrons and positrons emit photons 
by scattering off the microwave radiation. However, on a scattering length, 
they are strongly deflected in intergalactic magnetic field and thus the 
cascade radiation is spread over a large solid angle. This makes the 
contribution of the cascade radiation to the flux within the angle of 
observation negligible \cite{SJ}. For the energy range of interest,
$E_{\gamma} \geq 1\cdot 10^3$~GeV, the energy of radiating electrons is
$E_e \geq 2\cdot 10^4$~GeV, and on scattering length,
$l=1/\sigma_T n_{bb}$, the electron is deflected in magnetic field to the angle
larger than $0.01~rad$,
unless the magnetic field in intergalactic space is smaller than 
$2\cdot 10^{-13}~G$, which may appear unrealistic.

Let us consider now neutrino radiation. If the observed gamma radiation is 
produced by protons then the ratio of intrinsic neutrino to gamma ray flux
is 1.1 for $\gamma=2.1$ and  1.0 for $\gamma=2.3$.

Deep underground, the high-energy neutrino flux is accompanied by an 
equilibrium muon flux, i.e the ratio of these fluxes is depth independent.
This ratio is determined by the neutrino-nucleon interaction and the muon 
energy losses. For the calculations of muon fluxes displayed in Fig.\,~3,
we used the ratios from Ref.\cite{BBDGP},\cite{Beres}. 
For values of $\gamma$ around 2.1--2.3  
the muon spectrum becomes steeper for muon energies higher 
than $\sim 1$~TeV. This is due to the fact that for energies below 1 TeV
both the neutrino-nucleon cross section and muon pathlength grow 
approximately linearly  
with energy, while above this value they grow very slowly.

As a consequence, most of the muons crossing the detector have energies 
about $1$~TeV. The calculated flux corresponds to $\sim 10$ muons  
with energies higher than $1$~TeV crossing the 
area $1$~km$^2$ per year. They can be marginally detected by the future 
detectors like AMANDA. It should 
be realised that the number of visible neutrino sources could be much
larger than the number of TeV gamma sources because of the absence of 
absorption of neutrinos.

In conclusion, the observations of HEGRA taken at face value 
imply that the flux of extragalactic 
IR radiation is much lower than was predicted before. The combined results of 
HEGRA and CASA favour the explanation of a 50 TeV cutoff due to absorption 
on IR radiation. In this case the flux of IR radiation is almost 
precisely    
fixed. The flux at $E \sim 1-2$~TeV is within 
detectability power of, say, the Whipple telescope. The accompaning neutrino 
radiation can be marginally detected by a future $1$~km$^2$  detector like
an expanded AMANDA.\\*[1cm] 
{\bf Acknowledgements}\\*[1cm]
The authors are grateful to J. Cronin and H. Meyer for illuminating 
discussions. We thank also G. Navarra for giving us the results of 
the EASTOP detector before publication and for useful discussion.

This work was supported, in part, by the Human Capital and Mobility Program
of the European Economic Community under contract No. CHRX-CT93-0120 (DG 12 
COMA).
L.B. was supported by the Swedish Natural Science Research Council (NFR).
H.R.R. wishes to thank SISSA for hospitality when part of this work was
performed, and received partial support from EU grant ERBFMRXCT 960090.

\newpage
\newpage
\begin{figure}[ht]
\includegraphics[width=15cm]{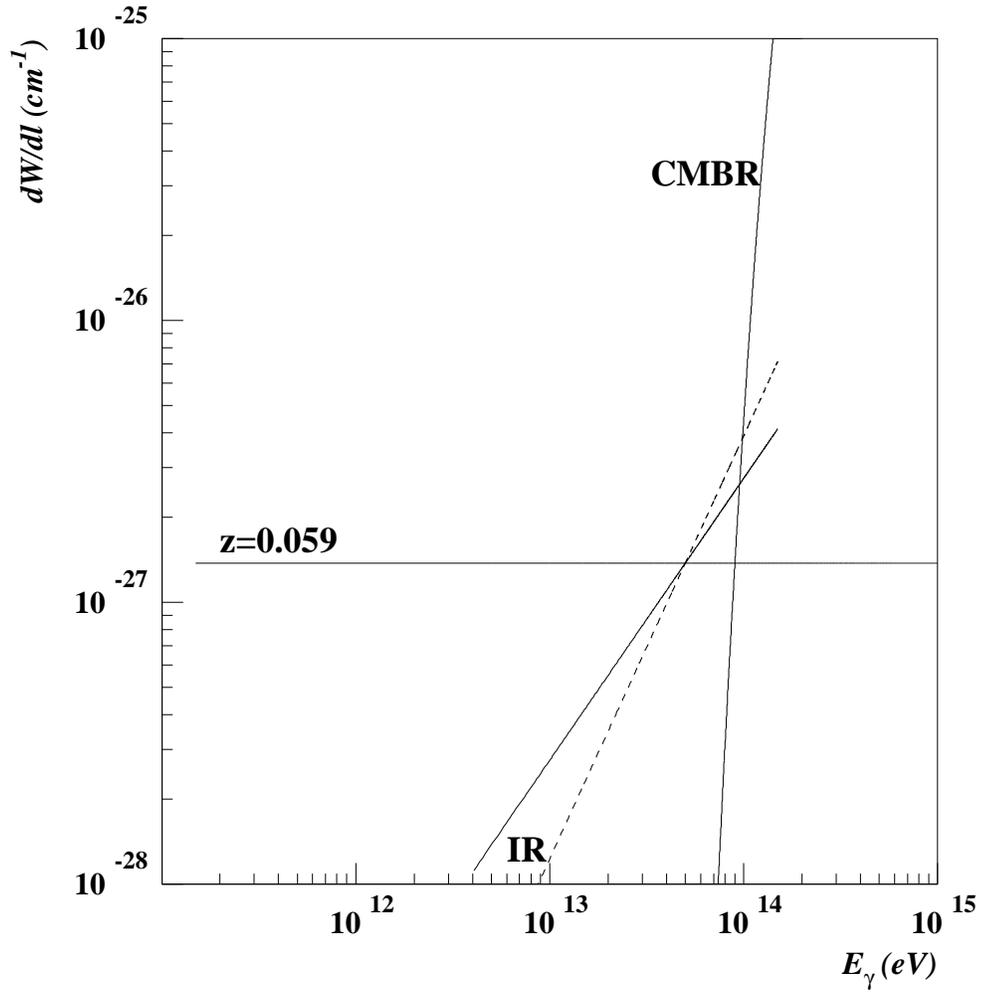}
\caption{Probability of photon absorption (inverse absorption length) on
microwave radiation (CMBR) and IR radiation in case of $\nu=2$ (solid curve
) and $\nu=2.5$ (dashed curve).
The curve labelled $z=0.059$ shows the inverse distance to the source 
0116+319.}
\label{fig:fig2}
\end{figure}
\newpage
\unitlength1cm
\begin{picture}(10,1)
 \epsfig{file={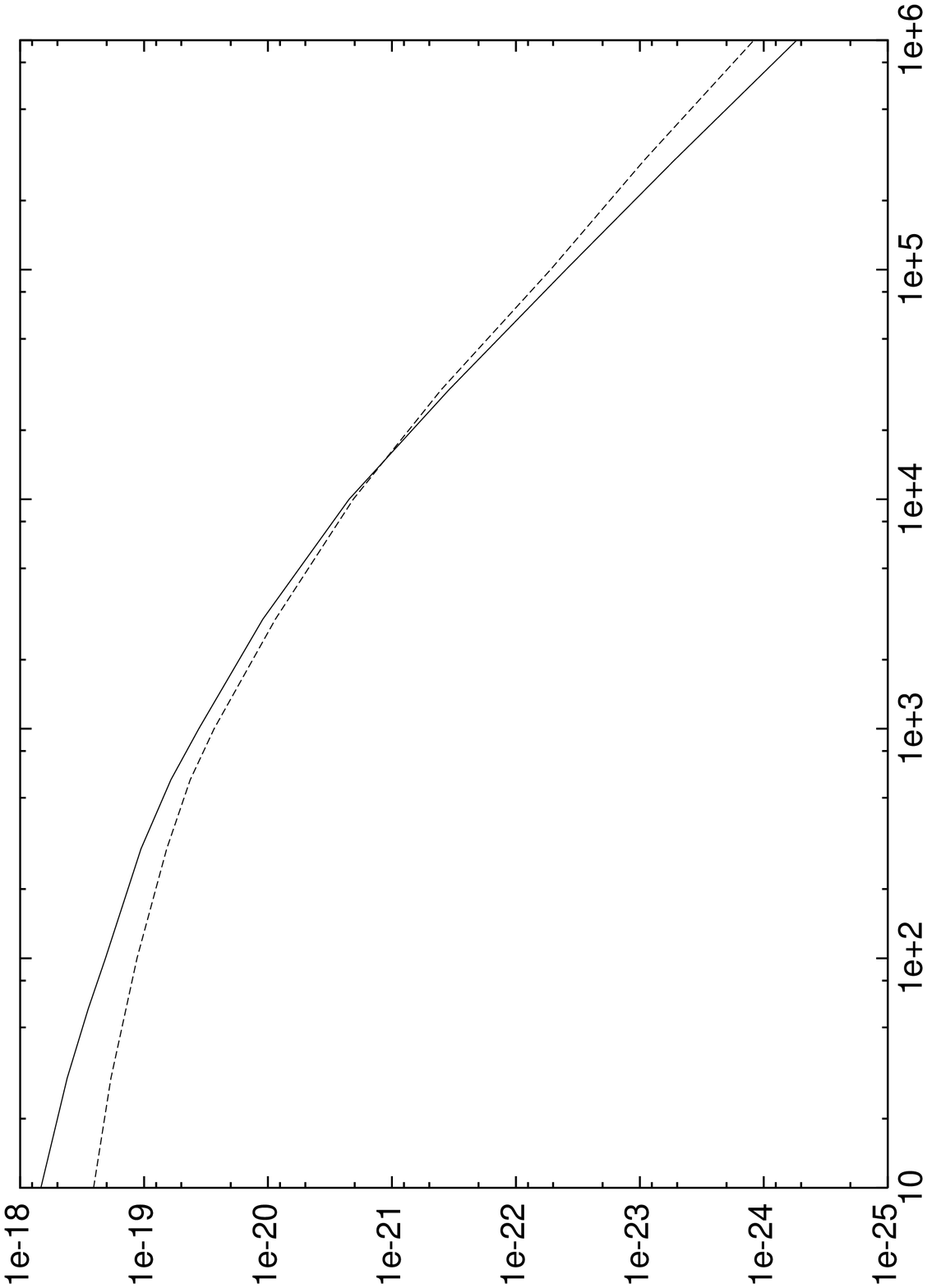}, height=10cm,angle=270}
 \put (-5.6,-7.8)
    {\footnotesize $E_\mu$ (GeV) }
 \put (-10.8,-4.5)
 {\begin{sideways}
     {\footnotesize $F_\mu$ (cm$^{-2}$s$^{-1}$GeV$^{-1}$) }
   \end{sideways}}
\end{picture}
\vskip9cm
\noindent
Figure 3: Deep underground equibrium muon flux produced by high energy 
neutrinos from 0116+319 source. The energy losses of muons are taken for 
water. The solid curve corresponds the neutrino spectrum index $\gamma=2.3$, 
and dashed curve - to $\gamma=2.1$ 
Flux of muons for the neutrino spectrum index $\gamma=2.3$ is 
higher at low energies than that for $\gamma=2.1$ because both fluxes are 
normalized by the HEGRA gamma ray flux at $E=50~TeV$. 

\end{document}